\documentclass[a4paper,11pt]{article}
\usepackage{pos}

\newcommand{\YY}{{Y}}

\usepackage{amsmath}
\newcommand{\ringring}[1]{%
  {% make an Ord atom
   \mathop{\kern0pt #1}\limits^{% set a box over the variable
     \vbox to-1.85ex{
       \kern-2ex % lower the ring accents
       \hbox to 0pt{\hss\normalfont\kern.1em \r{}\kern-.6em \r{}\hss}%
       \vss % fill
     }% end of \vbox
   }% end of the superscript
  }% end of \mathop
}

\title{The matrix edge of holography
}
\author[a,b]{Franz Ciceri}
\author[b,c]{Henning Samtleben}

\affiliation[a]{Laboratoire de Physique Théorique et Hautes Energies, Sorbonne Université, CNRS, F-75005 Paris, France}

\affiliation[b]{ENSL, CNRS, Laboratoire de physique, F-69342 Lyon, France}

\affiliation[c]{Institut Universitaire de France (IUF)}

\emailAdd{ciceri@lpthe.jussieu.fr}
\emailAdd{henning.samtleben@ens-lyon.fr}

\abstract{The IKKT matrix model arises at the extremal $p= -1$ limit of holographic dualities based on D$p$-brane geometries. We review the one-dimensional maximal supergravity that governs bulk fluctuations dual to the lowest BPS multiplet of gauge-invariant operators in the IKKT model. We present the Killing spinor equations and discuss their general half-supersymmetric solutions within the $\rm{SO}(3)\times \rm{SO}(7)$-invariant subsector. The explicit uplift of these solutions to Euclidean IIB supergravity in ten dimensions is provided.
}

\FullConference{Proceedings of the Corfu Summer Institute 2025 "School and Workshops on Elementary Particle Physics and Gravity" (CORFU2025)\\
27 April - 28 September, 2025\\
Corfu, Greece\\}

\begin{document}
\maketitle

\section{Introduction}
Holographic duality remains a central framework for exploring the non-perturbative dynamics of quantum field theories. Beyond the highly symmetric setting of conformal examples, the correspondence between type II strings on near-horizon D$p$-brane geometries and maximally supersymmetric Yang--Mills (SYM) theories in $p+1$ dimensions provides a rich arena for studying non-conformal holography. These systems retain sixteen supercharges while exhibiting scale-dependent dynamics in various dimensions \cite{Itzhaki:1998dd,Boonstra:1998mp}. In this contribution, we review recent progress \cite{Ciceri:2025maa,Ciceri:2025wpb} on a particularly intriguing limit of this correspondence, the extremal case $p=-1$. Here, the field-theory side reduces to the zero-dimensional IKKT matrix model, originally proposed as a non-perturbative definition of type IIB string theory \cite{Ishibashi:1996xs}. Despite extensive numerical investigations \cite{Krauth:1998xh,Ambjorn:2000dx,Anagnostopoulos:2022dak}, the holographic interpretation of this matrix integral has long remained largely unexplored. Recently, however, renewed interest has been driven by its maximally supersymmetric mass deformation, the polarized IKKT model \cite{Bonelli:2002mb}, which provides a tractable setting for concrete tests of `timeless holography'~\cite{Hartnoll:2024csr,Komatsu:2024bop,Komatsu:2024ydh,Hartnoll:2025ecj,Chou:2025rwy}.

D$p$-brane holographic dualities in higher dimensions share several key properties. The ten-dimensional dual geometry is conformal to the product of a round $S^{8-p}$ sphere and a $(p+2)$-dimensional anti-de Sitter space. In all cases, the $1/2$-BPS multiplets of single-trace operators in the SYM theory are in one-to-one correspondence with the fluctuations of ten-dimensional supergravity around the $S^{8-p}$ geometry \cite{Morales:2004xc}. In particular, the lowest BPS multiplet is dual to the lowest Kaluza--Klein supergravity multiplet in the bulk. On the gravity side, the dynamics of this multiplet is captured at the full non-linear level by a maximal gauged supergravity in $p+2$ dimensions with gauge group $\mathrm{SO}(9-p)$, arising from the consistent truncation of ten-dimensional supergravity on $S^{8-p}$. This provides a powerful and controlled framework for the quantitative analysis of such dualities, as developed in \cite{Bianchi:2001kw,Wiseman:2008qa,Kanitscheider:2008kd} and, for example, more recently explored in \cite{Bobev:2025idz}. Here, we present the gravitational construction for the $p=-1$ case, where the dual background is given by the $1/2$-supersymmetric D$(-1)$ instanton solution \cite{Gibbons:1995vg,Gubser:1996wt,Bergshoeff:1998ry,Ooguri:1998pf}. In this setting, the relevant gauged supergravity is a one-dimensional maximal supergravity, whose field content realizes the lowest BPS multiplet of single-trace operators of the IKKT matrix model.

The IKKT matrix model is defined by the action
\begin{equation}
S_{\text{\tiny{IKKT}}}=
-
\text{Tr}\Big[\tfrac14[X_a,X_b][X^a,X^b]-\tfrac12\bar{\Psi}\,\Gamma^a\,[X_a,\Psi]\Big]\,,
\label{eq:IKKTaction}
\end{equation}
where, $X_a$ and $\Psi^{\alpha}$ are bosonic and fermionic ${\mathfrak su}(N)$-valued matrices, respectively.
Vector indices $a,b=1, \dots, 10$, are contracted in Euclidean signature, and the $\Gamma_a$ denote the $\mathrm{SO}(10)$ $\Gamma$-matrices. The spinors $\Psi^{\alpha}$, $\alpha=1, \dots, 32$, are complex, and 
the charge-conjugated spinors are defined as 
\begin{equation}
    \bar \Psi_\alpha:=\Psi^\beta\mathcal C_{\alpha\beta}
    \,,\label{eq:CCIKKT}
\end{equation}
such that the Hermitian conjugate of $\Psi^\alpha$ does not appear in the (complex) action $S_{\text{\tiny{IKKT}}}$. 
In most formulae, we suppress the explicit spinor indices $\alpha$.
The charge conjugation matrix in (\ref{eq:CCIKKT}) is defined as $\mathcal C=-i\Gamma_{10}\Gamma_{*}$, with $\Gamma_{*}=-i\Gamma_{1}\Gamma_2\ldots\Gamma_{10}$. In addition, the spinors in (\ref{eq:IKKTaction})
satisfy the chirality condition
\begin{equation}
\Gamma_{*}\Psi=\Psi
\,.
\label{eq:chirality}
\end{equation}

In \cite{Ciceri:2025maa,Ciceri:2025wpb}, we constructed the bulk realization of the lowest BPS multiplet of ${\mathfrak su}(N)$-invariant single-trace operators in the IKKT matrix model. This multiplet combines the operators
\begin{align}
{\bf 54} \ : \quad &
{\cal O}^{ab} = {\rm Tr}[X^{a} X^{b}]-\tfrac1{10}\,\delta^{ab}\,{\rm Tr}[X^{c} X_{c}] 
\,,\nonumber\\
{\bf 144}_{s}\ : \quad &
{\cal O}^a = {\rm Tr}[X^a\,\Psi] - \tfrac19\, {\rm Tr}[X_b\,\Gamma^{ab} \Psi]
\,,\nonumber\\
{\bf 120} \ : \quad &
{\cal O}^{abc} = {\rm Tr}\left[X^{a} [X^{b},X^c]\right] -\tfrac18\, {\rm Tr}\left[\bar\Psi \Gamma^{abc} \Psi \right]
\,,
\label{eq:BPS1}
\end{align}
where the numbers count the ${\rm SO}(10)$ representations.
To complete the counting of degrees of freedom, one still needs to subtract
\begin{align}
{\bf 45} 
\oplus
{\bf 16}_{c}
 \,,
 \label{eq:BPS2}
\end{align}
accounting for the global ${\rm SO}(10)$ symmetry and supersymmetry, which in zero dimensions eliminate scalar and spinor degrees of freedom, respectively. In the holographic dual, both (\ref{eq:BPS1}) and (\ref{eq:BPS2}) are realized as bulk fields whose dynamics are described, at the fully nonlinear level, by a one-dimensional maximal supergravity. The latter correspond to gauge fields and gravitini which in the one-dimensional bulk theory couple as Lagrange multipliers.
In the following, we review the resulting one-dimensional maximal supergravity constructed in \cite{Ciceri:2025maa,Ciceri:2025wpb}. We present its 1/2-BPS solutions with $\rm{SO}(3)\times \rm{SO}(7)$ symmetry, and their explicit uplift to Euclidean IIB supergravity.

\section{Maximal supergravity in one dimension}

In this section, we review the maximal SO(10) gauged supergravity in one dimension that describes the bulk realization of the BPS multiplet (\ref{eq:BPS1}) and (\ref{eq:BPS2}). Its field content is given by
\begin{equation}
 {\rm bosonic:}\;\;   \{ {\rm e}, \phi, A_{[ij]}, T_{(ij)}, a_{[ijk]} \}\,,
    \qquad
    {\rm fermionic:}\;\;
    \{ \psi^\alpha, \lambda^\alpha, \chi^\alpha_{a} \}
    \,,
\end{equation}
where $i,j=1,\ldots,10$. The bosonic fields combine the einbein ${\rm e}$, dilaton $\phi$, and \textbf{45} $\mathrm{SO}(10)$ gauge fields $A_{ij}=-A_{ji}$ with \textbf{54} scalar fields parametrizing a symmetric ${\rm SL}(10)$ matrix $T_{ij}={V}_i{}^a {V}_{j}{}^b\delta_{ab}$, and \textbf{120} axions $a_{ijk}=a_{[ijk]}$\,. The fermions contain the gravitini $\psi$, dilatini $\lambda$ and matter fermions $\chi_a$, with the spinor conventions from (\ref{eq:CCIKKT}). The construction also features the dressed scalar fields
\begin{equation} \mathcal{T}^{ab}=\delta^{ij}\,V_i{}^aV_j{}^b\,,
\qquad
a_{abc}=(V^{-1})_a{}^i(V^{-1})_b{}^j(V^{-1})_c{}^k\,a_{ijk}\,.
\label{eq:axionSO}
\end{equation} 
In the following, contractions of 
$i,j,\ldots$ indices with the SO(10)-invariant tensor $\delta^{ij}$are left implicit.

To second order in the fermionic fields, 
the Lagrangian of the one-dimensional theory
can be organized as
\begin{equation}
\mathcal L_{\text{SUGRA}}=\mathcal L_{\text{0}}+\mathrm{g}\,\mathcal L_{\text{top}}+\mathrm g\,\mathcal L_{\text{Yukawa}}+\mathrm{g}^2\mathcal L_{\text{pot}}\,,\label{eq:LSUGRA}
\end{equation}
with the gauge coupling constant $\mathrm{g}$.
The first term depends on the gauge coupling constant and gauge fields only implicitly through the covariant derivatives and currents. Its explicit expression is given by
\begin{align}
\mathcal L_0=&\,10\,\mathrm{e}^{-1} \dot \phi^2-\mathrm{e}^{-1}P^{ab} P_{ab}
-\frac1{12}\,\mathrm{e}^{-1}e^{-2\phi}\,p_{abc}\,p^{abc}
+20\,\bar\lambda\,\mathcal D_t\lambda+2\,\bar\chi^a\,\mathcal D_t\chi_a
\nonumber\\&{}
-20\,\mathrm{e}^{-1}\bar\psi\,\Gamma_*\lambda\,\dot\phi
+2\,\mathrm{e}^{-1}\bar\psi\,\Gamma_{b}\chi_{a}\, p^{ab}-\frac{1}{2}\mathrm{e}^{-1}e^{-\phi}\,\bar\chi_{a}\,\Gamma_{bc}\,\psi\,p^{abc}
\nonumber\\
\,&{}
-\frac{1}{12}\,e^{-\phi}\,\bar\chi^{a}\,\Gamma_{bcd}\,\chi_a\,p^{bcd}
-e^{-\phi}\,\bar\chi_a\, \Gamma_b\, \chi_c\,p^{abc}-\frac{1}{6}\,\mathrm{e}^{-1}e^{-\phi}\,\bar\lambda\,\Gamma_{abc}\Gamma_*\,\psi\,p^{abc}
\nonumber\\[2mm]
\,&{}
-e^{-\phi}\,\bar\lambda\,\Gamma_{abc}\,\lambda\,p^{abc}-e^{-\phi}\,\bar \chi_c\,\Gamma_{ab}\Gamma_*\,\lambda\,p^{abc}\,,
\label{eq:L0}
\end{align}
with gauge covariant scalar currents defined as
\begin{align}
    & (V^{-1})^{a}{}^{i} D_t V_{i}{}^b=(V^{-1})^{a}{}^{i}  \left(\dot V_{i}{}^b-\mathrm g\,A_{ji}V_{j}{}^b\right)
=J^{ab} =: Q^{[ab]} + P^{(ab)}
\,,\\
&
     (V^{-1})_a{}^i(V^{-1})_b{}^j(V^{-1})_c{}^k\,
    D_t a_{ijk}=(V^{-1})_a{}^i(V^{-1})_b{}^j(V^{-1})_c{}^k\,\big( \dot a_{ijk}-3\,\mathrm{g}\,A_{l[i}\,a_{jk]l}\big)
= :p_{abc}
\,.\nonumber
\end{align}
We use dots to denote derivatives with respect to the one-dimensional coordinate $t$. Fermionic covariant derivatives are defined as
\begin{align}
\mathcal{D}_t \psi&=\dot\psi+\frac14 Q^{ab}\,(\Gamma_{ab})\,\psi\,,\nonumber\\
\mathcal D_t\chi_a&=\dot\chi_a+\frac14 Q^{bc}\,\Gamma_{bc}\,\chi_a+ Q^{ab}\,\chi_b\,,
\end{align} 
and analogously for $\lambda$. The first line of (\ref{eq:L0}) comprises the kinetic terms for the various fields, the remaining terms represent the Noether-type couplings between the fermions and the scalar currents. In one dimension, there is no Einstein-Hilbert term, rather gravity couples via the algebraic couplings of the einbein~${\rm e}$. Similarly, there is no Rarita-Schwinger term for the gravitino $\psi$, which appears as a Lagrange multiplier.

The second term in the Lagrangian \eqref{eq:LSUGRA} corresponds to a `topological' axion term in one dimension. It is independent of the einbein ${\rm e}$ and involves a single derivative,
\begin{align}
    \mathcal L_{\text{top}}=&\,\frac{\mathrm{i}}{1152}\,\varepsilon_{a_1a_2a_3a_4a_5a_6a_7a_8a_9 a_{10}}\,\mathcal{T}_{bc}\,a^{ba_1a_2}\,a^{ca_3a_4}\,a^{a_5a_6a_7}\,p^{a_8a_9a_{10}}\,.
    \label{eq:Ltop}
\end{align}
The third term in \eqref{eq:LSUGRA} represents Yukawa-like interactions between fermionic and scalar fields. It can be explicitly given as the following combinations of spinor bilinears,
\begin{equation}
\mathcal L_{\text{Yukawa}}=-40\,\bar\lambda\,B\,\psi-4\,\bar\chi_a\,C^a\,\psi+\mathrm{e}\,\bar\lambda\,E\,\lambda+\mathrm{e}\,\bar\lambda\,E^a\chi_a+\mathrm{e}\,\chi_a\,E^{ab}\chi_b\,,\label{eq:YukawaAnsatz}
\end{equation}
with the scalar dependent tensors $B\,,C^a,\,E\,,E^a\,,E^{ab}$ given by 
\begin{align}
B=&\,\frac{b}{10}+\frac{3}{80} b^{abc}\,\Gamma_{abc}+\frac{1}{160}b^{abcd}\,\Gamma_{abcd}
\,,\nonumber\\[2mm]
C^a=&-\frac12 b^{ab}\,\Gamma_b\Gamma_*-\left(\frac{7}{80}\,b_{bcd}-\frac{1}{6}\,b_{b,cd} \right)\tilde\Gamma^{ab,cd}\Gamma_*
+\frac{3}{40}b_{bcde} \,\tilde\Gamma^{ab,cde}\Gamma_*
\,,\nonumber\\[2mm]
E=&-21\,b \, \Gamma_*-7\,b^{abc}\,\Gamma_{abc}\Gamma_*-\frac{9}{8}b^{abcd}\,\Gamma_{abcd}\Gamma_* \,,\nonumber\\[2mm]
E^a=&-16\,b^{ab}\, \Gamma_b+\left(3\,b^{abc}-4\,b^{a,bc}\right)\,\Gamma_{bc}-2\,b^{abcd}\,\Gamma_{bcd}\,,\nonumber\\[2mm]
E^{ab}=&- \left( 4\,b^{ab}-\frac1{10}\,\delta^{ab}\,b\right)\,\Gamma_*-\frac14 \delta^{ab}\,b^{cde}\,\Gamma_{cde}\Gamma_* +\frac13 \left(3\,b^{acb}-4\,b^{c,ab}\right)\,\Gamma_{c} \Gamma_*\nonumber\\[1.5mm]
&\,\,+\frac{1}{16}\delta^{ab}\,b^{cdef}\,\Gamma_{cdef}\Gamma_* 
-\frac{3}{2}b^{abcd}\,\Gamma_{cd}\Gamma_*\,,\label{eq:YukawaAll} 
\end{align}
in terms of the irreducible $\mathrm{SO}(10)$ tensors, 
\begin{align}
b&=e^{4\phi}\,\mathcal{T}^a{}_a\,,\nonumber\\
b^{ab}&=\,e^{4\phi}\left(\mathcal{T}^{ab}-\tfrac{1}{10}\,\delta^{ab}\,\mathcal{T}^c{}_c\right)\,,\nonumber\\
b^{abc}&=\,e^{3\phi}\,\mathcal{T}^{d[a}\,a^{bc]}{}_d\,,\nonumber\\
b^{a,bc}&=\,e^{3\phi}\left(\mathcal{T}^{da}\,a^{bc}{}_d-\mathcal{T}^{d[b}\,a^{c]a}{}_d\right)\,,\nonumber\\
b^{abcd}&=\,e^{2\phi}\,\mathcal{T}^{ef}\,a_e{}^{[ab}\,a^{cd]}{}_f\,.\label{eq:babcde}
\end{align}
Here, we have introduced the gamma matrix combinations
\begin{equation}
\tilde\Gamma^{a_1a_2\,,\, a_3\ldots a_n}:=\delta^{a_1a_2}\,\Gamma^{a_3\ldots a_n}-\frac{1}{11-n}\Gamma^{a_1\ldots a_n}\,,
\label{eq:Gammacombi}
\,
\end{equation}
for $n>2$, which satisfy $\Gamma_{a_1}\tilde\Gamma^{a_1[a_2\,,\,a_3\ldots a_n]}=0$.

The last term in the Lagrangian \eqref{eq:LSUGRA} describes the scalar potential, which can be written compactly as
\begin{align}
    \mathcal{L}_{\text{pot}}=-\mathrm{e}\,V_{\text{pot}}=-\mathrm{e}\left(b^{ab}\,b_{ab}-\frac25\,b^2
    -\frac14\,b^{abc}\,b_{abc}+\frac29\,b^{a,bc}\,b_{a,bc}+\frac3{16}\,b^{abcd}\,b_{abcd}\right)\,,  \label{eq:Lpot}
\end{align}
in terms of the tensors (\ref{eq:babcde}). Explicitly, it takes the form
\begin{align}
V_{\text{pot}}=&\,\frac12\, e^{8\phi}\left(2\, \mathcal{T}^{ab}\,\mathcal{T}_{ab}-(\mathcal{T}^{a}{}_a)^2\right)-\frac14 \,e^{6\phi}
\left(2\,\mathcal{T}^{ac}\,\mathcal{T}^{bd}\,a_{abe}\,a_{cd}{}^{e}-\mathcal{T}^{ac}\,\mathcal{T}^{b}{}_{c}\,a_{ade}\,a_{b}{}^{de}\right)\nonumber\\
&\,+\frac3{16}\,e^{4\phi}\,\mathcal{T}^{ef}\,\mathcal{T}^{gh}\,a_e{}^{[ab}\,a^{cd]}{}_f\,a_{gab}\,a_{cdh}\,,
\label{eq:Vexp}
\end{align}
in terms of the dressed scalar fields (\ref{eq:axionSO}). It is a fourth-order polynomial in the axion fields.

The precise couplings of the Lagrangian \eqref{eq:LSUGRA} are fully fixed by imposing maximal supersymmetry, which we have checked to quadratic order in the fermions.
The supersymmetry transformation rules read
\begin{align}
\delta_\epsilon \mathrm{e}&=\bar\epsilon\, \psi\,,\nonumber\\[1mm]
\delta_\epsilon \phi&=\bar\epsilon\,\Gamma_*\lambda\,,\nonumber
\\[1mm]
\delta_\epsilon V_i{}^a&=V_i{}_b\,\,\bar\epsilon\,\Gamma^{(a}\,\chi^{b)}\,,\nonumber\\[1mm]
\delta_\epsilon a_{ijk}&=-e^\phi\,V_{[i}{}^a V_{j}{}^{b} V_{k]}{}^{c}\left(
3\,\bar\epsilon\,\Gamma_{ab}\,\chi_{c}
+\bar\epsilon\,\Gamma_{abc}\Gamma_*\,\lambda\right)\,,\nonumber\\[1mm]
\delta_\epsilon A_{ij}=&\,e^{4\phi}\,V_{[i}{}^a V_{j]}{}^{b}\left(\bar\epsilon \,\Gamma_{ab}\Gamma_*\,\psi+4\,\mathrm{e}\,\bar\epsilon\,\Gamma_{ab}\,\lambda-2\,\mathrm{e}\,\bar\epsilon\,\Gamma_{a}\Gamma_*\,\chi_b\right)\nonumber\\[1mm]
&\,+e^{3\phi}\,V_{[i}{}^b V_{j]}{}^{c}\,a_{abc}\left(\bar\epsilon\,\Gamma^a\Gamma_*\,\psi+3\,\mathrm{e}\,\bar\epsilon\,\Gamma^a\,\lambda+\mathrm{e}\,\bar\epsilon\,\Gamma_*\,\chi_a\right)\,,
\label{eq:SUSYbos}
\end{align}
for the bosonic fields, and 
\begin{align}
\delta_\epsilon \psi&= \mathcal D_t\,\epsilon
+\frac1{24}e^{-\phi}\,p^{abc}\,\Gamma_{abc}\,\epsilon \,
+\mathrm{e} \,\mathrm{g}\,A\,\epsilon\,,\nonumber\\[1mm]
\delta_\epsilon\lambda&=\frac{1}{2}\mathrm{e}^{-1}\dot\phi\,\Gamma_{*}\,\epsilon
+\frac{1}{240}\mathrm{e}^{-1}e^{-\phi}\,p^{abc}\,\Gamma_{abc}\Gamma_*\,\epsilon 
+\mathrm g\,B\,\epsilon\,,\nonumber\\[1mm]
\delta_\epsilon\chi^{a}&=\frac{1}{2}\mathrm{e}^{-1}P^{ab}\,\Gamma_{b}\,\epsilon
+\frac{7}{80}\mathrm{e}^{-1}\,e^{-\phi}\,p_{bcd}\,\tilde\Gamma^{ab,cd}\,\epsilon
+\mathrm g\,C^a\,\epsilon\,,
\label{eq:SUSYfer}
\end{align}
for the fermionic fields. The fermion shifts in (\ref{eq:SUSYfer}) are given by the Yukawa tensors $B$ and $C^a$ from (\ref{eq:YukawaAll}) together with
\begin{equation}
A=-\frac14\,b\,\Gamma_*+\frac18 b^{abc}\,\Gamma_{abc}\Gamma_* -\frac{1}{32}b^{abcd}\,\Gamma_{abcd}\Gamma_* 
\,,
\end{equation}
for the gravitino variation.
From the variation of the fermionic fields (\ref{eq:SUSYfer}) one can in particular read off the Killing spinor equations of the theory. This completes the description of the one-dimensional maximal supergravity.

 Let us note that unlike in higher dimensions, where gauged supergravities are in general constructed as deformations of the toroidally reduced theory with its characteristic symmetry enhancement, the one-dimensional maximal supergravity (\ref{eq:LSUGRA}) has been constructed directly as a non-abelian gauged theory in one dimension. The relation between its ungauged limit $\mathrm{g}\rightarrow0$ and toroidally reduced supergravity remains somewhat mysterious and calls for further exploration.

 Nonetheless, we expect the gauged theory (\ref{eq:LSUGRA}) to have a higher-dimensional origin and to arise from consistent truncation of Euclidean IIB supergravity on a nine-sphere $S^9$, describing the full non-linear couplings of the lowest Kaluza-Klein fluctuations. For the bosonic theory with vanishing axions $a_{ijk}=0$, this has been shown in \cite{Ciceri:2025maa}, based on previous work in higher dimensions \cite{deWit:1986oxb,Nastase:2000tu,Cvetic:2000dm,Ciceri:2023bul}. Explicitly, this gives rise to the uplift formulae (in the Einstein frame)
\begin{align}
d s_{10}^2 =\,&  e^{9\phi}\,\Delta\, 
\mathrm{e}^2\,dt^2 
+ \mathrm{g}^{-2}\, e^{\phi} \, ({T}^{-1})^{ij}\, 
{ D}\mu_i\, { D}\mu_j
\,,\nonumber\\[1ex]
e^ {\Phi} =\,&  
e^{-4\phi}\,\Delta^{-1}
\,,\nonumber\\
{\cal X}=&-\frac1{2\,\mathrm{e}\mathrm{g}}
\left(
(T^{-1})^{ij}\, D_t T_{kj}\,\mu_{i}\mu_{k}-\dot\phi\right)\,,\label{eq:Ansatz3}
\end{align}
with the embedding coordinates of the unit $S^9$ sphere denoted as $\mu_i$, with $i=1,\ldots,10$, and $\mu_i\mu_i$=1. $\mathrm{SO}(10)$ gauge covariant derivatives act as 
\begin{align}
{ D} \mu_i =\,& d\mu_i - \mathrm{g}\, A_{ji}\,\mu_j\,dt\,,\\
{ D}_t  T_{ij} =\,&
  \dot{T}_{ij} -2\, \mathrm{g}\,A_{k(i}  T_{j)k}\,,
\label{eq:covD}
\end{align}
and $\Delta=T_{ij}\,\mu_i\mu_j$.
If the fields $\{\mathrm{e}, \phi, T_{ij}\}$ satisfy the field equations of (\ref{eq:LSUGRA}), the ten-dimensional fields defined by (\ref{eq:Ansatz3}) satisfy the field equations of the Lagrangian
\begin{equation}
{\cal L}_{\rm IIB} =
 |E|\Big(R - \tfrac12 \partial_\mu \Phi \,\partial^\mu\Phi  
+\tfrac12 e^{2\Phi}\, \partial_\mu {\cal X} \,\partial^\mu {\cal X} \Big) \,,\label{eq:LEu}
\end{equation}
which is the subsector of Euclidean IIB supergravity with vanishing p$>0$-forms. In particular, the ${\rm SO}(10)$-invariant solution 
\begin{equation}
  T_{ij}=\delta_{ij}\,,\quad  
  a_{ijk}=0\,,\quad
  A_{ij}=0\,,\quad
e^\phi=\mathrm{g}^2\,t^2\,,\quad
{\rm e} =\mathrm{g}^{-9}\,t^{-9}\,,
\label{eq:D-1}
\end{equation}
of (\ref{eq:LSUGRA}) uplifts to the 1/2-supersymmetric D$(-1)$ instanton solution 
\cite{Gibbons:1995vg,Gubser:1996wt,Bergshoeff:1998ry,Ooguri:1998pf}.

Switching on the axion fields $a_{ijk}$ not only modifies the uplift formulae (\ref{eq:Ansatz3}) for the ten-dimensional metric and axio-dilaton, but also generically induces nontrivial IIB supergravity p-forms. In the next section, we will extend the uplift formulae to the presence of one axion field, as allowed by the 1/2-BPS solutions with $\rm{SO}(3)\times \rm{SO}(7)$ symmetry.  
A systematic derivation of the general uplift formulae for the full Lagrangian (\ref{eq:LSUGRA}) would require its embedding into some suitable exceptional field theory, as has been achieved in other dimensions~\cite{Baguet:2015sma,Bossard:2022wvi,Bossard:2023jid}.

\section{BPS solutions}

\subsection{$\rm{SO}(3)\times \rm{SO}(7)$-invariant subsector and 1/2-BPS solutions}\label{sec:BPSwithaxion}
The general supersymmetric solution with vanishing axion fields has been given in \cite{Ciceri:2025maa} and is parametrized by 10 real constants.
Here, we consider supersymmetric solutions in the presence of non-vanishing axion fields, however preserving an 
\begin{equation}
  {\rm SO}(3)\times {\rm SO}(7)\subset   {\rm SO}(10) \,,
\end{equation}
subgroup of the original gauge group.
This smaller symmetry group allows for a non-trivial scalar matrix $T_{ij}$ of the form
\begin{equation}
    T_{ij} = {\rm diag}\big( 
    X^{7/10},X^{7/10},X^{7/10},
X^{-3/10},
X^{-3/10},
X^{-3/10},
X^{-3/10},
X^{-3/10},
X^{-3/10},
X^{-3/10}\big)
\,,
\label{eq:trunc1}
\end{equation}
as well as a non-vanishing dilaton and one axion field $a_{ijk}$ which we parametrize as
\begin{equation}
    e^{\phi}  =  e^{\eta/7} \,X^{-1/20}\,,\qquad
    a_{123}=
    e^{\eta/7} \, \YY\,,
    \label{eq:trunc2}
\end{equation}
respectively. On the other hand, all vector fields are set to zero. In this truncation, the Lagrangian (\ref{eq:LSUGRA}) reduces to 
\begin{align}
    {\cal L} =& -\frac12\,\mathrm{e}^{-1}\,\frac{\dot{X}^2+\dot{\YY}^2}{X^2}
    -\frac17\,\mathrm{e}^{-1}\,\frac{X\dot{X}+\YY\dot{\YY}}{X^2}\,\dot{\eta}
    +\frac{10}{49}\,\mathrm{e}^{-1}\,\Big(1-\frac{\YY^2}{20X^2}\Big)\,\dot\eta^2
    \nonumber\\
    &\;
    + \frac12\,\mathrm{g}^2\,\mathrm{e}\,e^{8\eta/7}\,
    \frac{35+42\,X+3\,X^2+3\,\YY^2}{X}
    \,.\label{eq:LXY}
\end{align}
Its equations of motion
are implied by the BPS equations 
\begin{equation}
\delta_\epsilon\psi\stackrel{!}{=}0\,,
\qquad\delta_\epsilon\lambda\stackrel{!}{=}0\,,
\qquad\delta_\epsilon \chi^a \stackrel{!}{=} 0\,,
\label{eq:KS1}
\end{equation}
from (\ref{eq:SUSYfer}).
In the truncation (\ref{eq:trunc1}), (\ref{eq:trunc2}), they take the form
\begin{align}
\frac{d\YY}{dX} \ = \ &  - \frac{\YY \left(7 X^2+\YY^2+8 X+9\right)}{2 X \left(2 (1-X) (X+3)+\YY^2\right)}\,,
   \nonumber\\
\frac{d\eta}{dX} \ = \ & \frac{7 \left((X+3)^2+\YY^2\right)}{2 X \left(2 (1-X) (X+3)+\YY^2\right)}
\,,
\label{eq:BPS}
\end{align}
together with
\begin{equation}
     \frac{dX}{dt}=
   \mathrm{g}\, \mathrm{e}\,e^{\frac{4 \eta}{7}} \frac{\sqrt X
   \left(2(X-1)(X+3)-\YY^2\right)}{\sqrt{(3+X)^2+\YY^2} }
   \,.
   \label{eq:BPS-B2}
\end{equation}
These first order differential equations are the Euclidean analogue of the $\mathrm{SO}(2,1)\times \mathrm{SO}(6-p)$ BPS equations for spherical D$p$-branes \cite{Bobev:2018ugk,Bobev:2024gqg}, extrapolated to $p=-1$, as discussed in \cite{Ciceri:2025wpb}. 

In addition, the equations (\ref{eq:KS1}) imply  the projection condition on the Killing spinor
\begin{equation}
\mathbb{P}\,\epsilon=0\,,\qquad   \mathbb{P}
    \equiv \frac12\,(\mathbb{I}-\Gamma_*)
   \Big(\mathbb{I}+
    \frac{63\, \mathrm{g}\,
    e^{4\eta/7}X^{1/2}\,\YY+7\,\dot\YY+\dot\eta\,\YY}{28\,\mathrm{g}\,e^{4\eta/7}\, \big(7+3 \,X\big)\,  X^{1/2}-20\,X\,\dot\eta+7\,\dot X}\,
    \Gamma_{123}\Big)\,,
\label{eq:projector}
\end{equation}
where $\mathbb{P}^2=\mathbb{P}$ and ${\rm tr}\,\mathbb{P} = 16$ show that solutions of (\ref{eq:BPS}) and (\ref{eq:BPS-B2}) preserve half of all supersymmetries. 
Let us note, that the BPS equations (\ref{eq:BPS}) imply that
\begin{equation}
    \frac{d}{dX}\left(
   8\mathrm{g}^2\, e^{\eta/7}\,\YY\,\big((1-X)(3+X)^2-X\YY^2\big)^{-1/2}
    \right) = 0\,.
\end{equation}
We will denote this constant as
\begin{equation}
   \mu = 8\mathrm{g}^2\, e^{\eta/7}\,\YY\,\big((1-X)(3+X)^2-X\YY^2\big)^{-1/2} {}\,.
   \label{eq:mu}
\end{equation}
We observe that the unique $\mathrm{SO}(10)$-invariant solution (\ref{eq:D-1}), which uplifts to the near-horizon limit of coincident D$(-1)$ instantons, takes the form
\begin{equation}
X=1\,,\quad
\YY=0\,,\quad e^\eta=\mathrm{g}^{14}\,t^{14}\,,\quad
\mathrm{e}=\mathrm{g}^{-9}\,t^{-9}\,.\label{eq:1dInstanton}
\end{equation}
Keeping ${Y}=0$ while turning on a nontrivial profile for $X(t)$ leads to an $\mathrm{SO}(3)\times \mathrm{SO}(7)$-invariant distribution of instantons in flat ten-dimensional space \cite{Ciceri:2025wpb}, dual to a vev deformation of the IKKT matrix model. In contrast, the solutions relevant for describing backreacted IIB geometries dual to polarized IKKT vacua should involve non-trivial profiles for both scalars $X(t)$ and $\YY(t)$.
\subsection{Uplift to Euclidean IIB supergravity}\label{sec:Upliftaxion}
In order to find the uplift of the above 1/2-BPS solutions, we consider an ansatz inspired by those of \cite{Bobev:2024gqg}. The final uplift to Euclidean IIB supergravity (in the Einstein frame) takes the form\footnote{We emphasize that this uplift only describes the 1/2-BPS solutions. It is not \textit{a priori} suitable to uplift solutions of (\ref{eq:LXY}) with less supersymmetries.}
\begin{align}\label{eq:UA}
    ds_{10}^2&=\frac{e^{\eta/7}}{\mathrm{g}^2\,\mathcal P^{1/4}\,\mathcal Q^{3/4}}\,\Big(\mathrm{g}^2\,\mathrm{e}^2\,e^{8\eta/7}dt^2+d\theta^2
    +\mathcal P\,\text{cos}^2\theta\,d\Omega_2^2+\mathcal Q\,\mathrm{sin}^2\theta\,d\Omega_6^2\Big)\,,\nonumber\\[2ex]
    e^\Phi&=e^{-4\eta/7}\sqrt{\mathcal P\mathcal Q}\,,\nonumber\\[1ex]
    {\cal X} &=-\frac{e^{4\eta/7}X^{1/2}}
{\sqrt{(3+X)^2+\YY^2}}\,
\Big(\frac{1}{\mathcal P}+\frac{3}{\mathcal Q}+\frac{Y^2}{2X^2}\big(\text{cos}^2\theta-\frac14\big)\Big)\,,\nonumber\\[1ex]
    B_{(2)}&= -e^{- \eta/7}\frac{\YY \mathcal P}{\mathrm{g}^2X}\,\mathrm{cos}^3\theta\,\mathrm{vol}_2\,,
\nonumber\\[1ex]
C_{(2)} &=
\frac{e^{3\eta/7}\,\YY\,\mathcal P\,{\rm cos}^3\theta}{\mathrm{g}^2\,X^{1/2}\,\sqrt{(3+X)^2+\YY^2}}\,
\Big(\frac{5}{2\mathcal Q}+\frac{1}{6 X\mathcal P}-\frac{Y^2}{8X^2}\Big)\,\text{vol}_2
\,,
\end{align}
with the angle $\theta\in[0,\frac{\pi}{2}]$ and the functions
\begin{align}
\mathcal P := \ &\frac{X}{X\mathrm{sin}^2\theta+(X^2+\YY^2)\,\mathrm{cos}^2\theta}\,,\qquad
\mathcal Q:=\frac{X}{\mathrm{sin}^2\theta+X\mathrm{cos}^2\theta}\,,
\end{align}
and where $\mathrm{vol}_n$ and $d\Omega^2_n$ denote the volume form and the metric on the unit sphere $S^n$, respectively. Note that the RR 4-form is set to zero.

Using the BPS equations (\ref{eq:BPS}) for the one-dimensional fields $X$, $\YY$ and $\eta$, we have verified that the uplift (\ref{eq:UA}) satisfies the equations of motion deriving from the Euclidean IIB supergravity Lagrangian\footnote{We follow the conventions of \cite{Komatsu:2024bop} for the Euclidean IIB theory, except that we treat $\mathcal X$ and $C_{(2)}$ as real, with corresponding kinetic terms of opposite sign.} 
\begin{equation}
    \mathcal L_{\text{IIB}}=\frac{1}{2\kappa_{10}^2}\int d^{10}x\Big(R-\frac12(\partial\Phi)^2+\frac12e^{2\Phi}(\partial\mathcal X)^2-\frac{1}{12}e^{-\Phi}(H_{(3)})^2+\frac{1}{12}e^{\Phi}(F_{(3)})^2\Big)\,,
\end{equation}
where the NS-NS and RR field strengths are defined as
\begin{align}
H_{(3)}=dB_{(2)}\,,\qquad F_{(3)}=dC_{(2)}-\mathcal X\,H_{(3)}\,.
\end{align}
The field equations for the axion $\mathcal X$ and the RR 2-form can be written as
\begin{align}
    F_{(7)}=e^\Phi\star F_{(3)}\,,\qquad F_{(9)}=e^{2\Phi}\star d\mathcal X+B_{(2)}\wedge F_{(7)}\,,
\end{align}
in terms of closed field strengths $F_{(7)}$ and $F_{(9)}$ for the dual RR 6- and 8-form. The ans\"atze for the latter can be computed from (\ref{eq:UA}) and read 
\begin{align}
 C_{(6)}&=6\,e^{\eta/7}\,\frac{\YY {\mathcal Q}}{\mathrm{g}^6X}\,\mathrm{sin}^7\theta\,\mathrm{vol}_6\,,\\
C_{(8)}&=-\mathrm{g}^{-8}\,\Big(\omega(\theta)+{\mathcal P}\,\mathrm{cos}\theta\,\mathrm{sin}^7\theta
+\frac{{\mathcal P}{\mathcal Q}\,\YY^2}{X^2}\,
\mathrm{cos}^3\theta\,\mathrm{sin}^7\theta
\Big)\,\mathrm{vol}_2\wedge\mathrm{vol}_6\,,   
\end{align}
The function $\omega(\theta)$ is defined by 
\begin{equation}
    \frac{d\omega}{d\theta}= \mathrm{sin}^6\theta\,.
\end{equation}
This ensures that when $X=1,\YY=0$, which implies ${\mathcal P}=\mathcal Q=1$, the 9-form flux $F_{(9)}=dC_{(8)}$ becomes proportional to $\mathrm{vol}_9$. This limit reproduces the SO(10)-invariant solution. Indeed, substituting (\ref{eq:1dInstanton}) into the uplift (\ref{eq:UA}) leads to 
\begin{align}
ds_{10}^2&=dt^2+t^2\big( d\theta^2+\text{cos}^2\theta\,d\hat\mu_I d\hat\mu_I+\text{sin}^2\theta \,d\tilde\mu_A d\tilde\mu_A\big)\,,\nonumber\\[1mm]
e^{\Phi}&=\mathrm{g}^{-8}t^{-8}=-\mathcal X^{-1}\,,\qquad B_{(2)}=0=C_{(2)}\,,
\label{eq:upliftD-1}
\end{align}
where $\hat\mu_{I}$ and $\tilde\mu_{A}$ respectively denote the embedding coordinates of the unit $S^2$ and $S^6$, and satisfy $\hat\mu_{I}\hat\mu_{I}=1=\tilde\mu_{A}\tilde\mu_{A}$, with $I=1,2,3$ and $A=4,\ldots,10$.
This recovers the
D$(-1)$ instanton solution of
\cite{Gibbons:1995vg,Gubser:1996wt,Bergshoeff:1998ry,Ooguri:1998pf}.
In particular, the metric (\ref{eq:upliftD-1}) describes Euclidean flat space. This is easily seen by switching back to the $S^9$ embedding coordinates $\mu_i$ of (\ref{eq:Ansatz3}), with $i=\{I\,,A\}$, via 
\begin{align}
\mu_{I}=\text{cos}\theta \,\hat\mu_{I}\,,\qquad\mu_{A}=\text{sin}\theta\,\tilde\mu_{A}\,.\end{align}

\subsection{Electrostatic potential formulation}\label{sec:Electrostatic}
It has been shown in \cite{Komatsu:2024bop}, based on the Lorentzian construction in \cite{DHoker:2016ujz}, that the general 1/2-BPS solution of Euclidean IIB supergravity with ${\rm SO}(3)\times {\rm SO}(7)$ symmetry can be brought into the form
\begin{align}
    ds_{10}^2 &= \frac{8}{\mu^{\frac{5}{2}}}\left( \frac{1}{3^3}\frac{ \Lambda \mathring{V}}{ (-V'') }\right)^{1/4}\left[ \frac{(-V'')}{ \mathring{V}} (d\rho^2+dz^2) +  \frac{\rho\, (-V'') \mathring{V}}{\Lambda} d\Omega_2^2 +3 \rho\,  d\Omega_6^2\right], \nonumber\\[1mm]
    e^{\Phi} &=  -\mu^3\frac{3 \mathring{V}+\rho V'' }{\rho \sqrt{3 \Lambda \mathring{V}(-V'')}}\,,\nonumber \\[1mm]
     \mathcal X &= -\mu^{-3}\frac{3 \mathring{V}(V'+\rho \mathring{V}')+\rho V' V''}{3 \mathring{V} + \rho V''}\,, \nonumber\\[1mm]
    B_{(2)} &= -\frac{8}{3\mu} \left(z - \frac{\rho \mathring{V} \mathring{V}'}{\Lambda} \right) \text{vol}_2 \nonumber \\[1mm]
   C_{(2)} &=  \frac{8}{3\mu^4}\left(V - \rho \frac{\mathring{V}}{\Lambda} \left(V' \mathring{V}'+3 \mathring{V}(-V'') \right) \right)\text{vol}_2\,, \label{eq:solutioninV}
\end{align}
with
\begin{equation}
 \Lambda := 3 \mathring{V}V''+\rho(\mathring{V}'^2+V''^2)
 \,,
\qquad\mathring{V}:=\partial_\rho V\,,\qquad 
 V':=\partial_z V\,,
\end{equation}
and the potential $V(\rho,z)$ satisfying the Laplace equation in a four-dimensional axially
symmetric system, 
\begin{equation}
    V'' + \ringring{V} + \frac{2}{\rho} \mathring{V} = 0\,,
\end{equation}
where $\rho$ is the radial coordinate and $z$ is the vertical direction.\footnote{The other two coordinates are not part of the ten-dimensional space (\ref{eq:solutioninV}).} When certain regularity and positivity conditions are imposed on the metric and dilaton, $V(z,\rho$) can be interpreted as the electrostatic potential associated to a distribution of conducting balls. As explained in detail in \cite{Komatsu:2024bop, Komatsu:2024ydh}, different ball configurations correspond to backreacted IIB geometries dual to different vacua of the polarized IKKT matrix model. 

For the uplifted 1/2-BPS solutions (\ref{eq:UA}),
we can achieve this form by the change of coordinates
\begin{equation}
    \rho = \frac{\mu\,e^{-\eta/7}\,\sqrt{X}\sqrt{(3+X)^2+\YY^2}}{8\,\mathrm{g}^2\,\YY}\,{\rm sin}\,\theta\,,\qquad
    z=  \frac{\mu\,e^{-\eta/7}\,(3+X)}{8\,\mathrm{g}^2\,\YY}\,{\rm cos}\,\theta
    \,,
\end{equation}
and a potential given by
\begin{equation}
    V = \frac{ \mu^4 \cos \theta \, e^{3 \eta/7} \,\YY\, \big(2 (X+1) \cos (2 \theta
   )+X-1\big)}{64\, \mathrm{g}^2\, X^{3/2} \sqrt{(3+X)^2+\YY^2}} \,.
   \label{eq:V}
\end{equation}
Equivalence of (\ref{eq:solutioninV}) and (\ref{eq:UA}) for this choice of $V$ can be shown straightforwardly, using the one-dimensional BPS equations (\ref{eq:BPS}). The parameter $\mu$ in (\ref{eq:solutioninV}) coincides with the integration constant from (\ref{eq:mu}).

\section{Conclusions}

We have reviewed the one-dimensional maximal supergravity which realizes the field content of the lowest BPS multiplet of gauge-invariant operators of the dual IKKT model. We presented the Killing spinor equations and discussed their solutions in the $\rm{SO}(3)\times \rm{SO}(7)$-invariant subsector. Finally, we gave the explicit uplift of the 1/2-BPS solutions to Euclidean IIB supergravity in ten dimensions. Together, this sets the stage for the holographic computation of correlation functions in the (polarized) IKKT model. In particular, the holographic dictionary identified in \cite{Ciceri:2025maa,Ciceri:2025wpb}, which relates the gauge-invariant operators in (\ref{eq:BPS1}) to the one-dimensional supergravity fields, implies that the scalars $X$ and $Y$ are dual to the operators
\begin{align}
&X\longleftrightarrow \mathcal O_X\sim \text{Tr}\,\Big[(X_1)^2+(X_2)^2+(X_3)^2\Big]-\frac{3}{10}\,\text{Tr}\Big[X^a X_a\Big]\,,\nonumber\\
&Y\longleftrightarrow \mathcal O_Y\sim \text{Tr}\,\Big[X_1[X_2,X_3]\Big]-\frac{1}{8}\,\text{Tr}\Big[\bar\Psi\,\Gamma_{123}\,\Psi \Big]\,.
\end{align}
%with $u$ only running over $1,2,3$.

\subsection*{Acknowledgements}
We would like to thank Nikolay Bobev, Adrien Martina, Jo\~ao Penedones, Antoine Vuignier, and Xiang Zhao for useful discussions. FC thanks the organizers of `Matrix Model for Superstring/M-theory' at the YITP, Kyōto for an enjoyable and stimulating workshop. HS thanks the organizers of the Corfu Summer Institute 2025 for a very enjoyable workshop. The research of FC is funded by the Deutsche Forschungsgemeinschaft (DFG, German Research Foundation) – Project number: 521509185.

\newpage

\bibliographystyle{utphys}
\bibliography{refs}

@article{Komatsu:2024ydh,
	archiveprefix = {arXiv},
	author = {Komatsu, Shota and Martina, Adrien and Penedones, Joao and Vuignier, Antoine and Zhao, Xiang},
	doi = {10.1007/JHEP12(2025)030},
	eprint = {2411.18678},
	journal = {JHEP},
	pages = {030},
	primaryclass = {hep-th},
	title = {Einstein gravity from a matrix integral. {P}art {II}},
	volume = {12},
	year = {2025}}

@article{Komatsu:2024bop,
	archiveprefix = {arXiv},
	author = {Komatsu, Shota and Martina, Adrien and Penedones, Jo{\~a}o and Vuignier, Antoine and Zhao, Xiang},
	doi = {10.1007/JHEP12(2025)029},
	eprint = {2410.18173},
	journal = {JHEP},
	pages = {029},
	primaryclass = {hep-th},
	title = {Einstein gravity from a matrix integral. {P}art {I}},
	volume = {12},
	year = {2025}}

@article{DHoker:2016ujz,
	archiveprefix = {arXiv},
	author = {D'Hoker, Eric and Gutperle, Michael and Karch, Andreas and Uhlemann, Christoph F.},
	doi = {10.1007/JHEP08(2016)046},
	eprint = {1606.01254},
	journal = {JHEP},
	pages = {046},
	primaryclass = {hep-th},
	title = {Warped {${\rm AdS}_6\times S^2$} in type {IIB} supergravity {I}: {L}ocal solutions},
	volume = {08},
	year = {2016}}

@article{Ciceri:2025wpb,
	archiveprefix = {arXiv},
	author = {Ciceri, Franz and Samtleben, Henning},
	doi = {10.1103/gmhl-mmg5},
	eprint = {2511.23111},
	journal = {Phys. Rev. D},
	number = {4},
	pages = {046001},
	primaryclass = {hep-th},
	title = {Supergravity dual for {I}shibashi-{K}awai-{K}itazawa-{T}suchiya holography},
	volume = {113},
	year = {2026}}

@article{Anagnostopoulos:2022dak,
	archiveprefix = {arXiv},
	author = {Anagnostopoulos, Konstantinos N. and Azuma, Takehiro and Hatakeyama, Kohta and Hirasawa, Mitsuaki and Ito, Yuta and Nishimura, Jun and Papadoudis, Stratos Kovalkov and Tsuchiya, Asato},
	doi = {10.1140/epjs/s11734-023-00849-x},
	eprint = {2210.17537},
	journal = {Eur. Phys. J. ST},
	number = {23-24},
	pages = {3681--3695},
	primaryclass = {hep-th},
	reportnumber = {KEK-TH-2470},
	title = {Progress in the numerical studies of the type {IIB} matrix model},
	volume = {232},
	year = {2023}}

@article{Chou:2025rwy,
	archiveprefix = {arXiv},
	author = {Chou, Chien-Yu and Nishimura, Jun and Wang, Cheng-Tsung},
	doi = {10.1103/y1rm-n85b},
	eprint = {2507.18472},
	journal = {Phys. Rev. Lett.},
	number = {22},
	pages = {221601},
	primaryclass = {hep-th},
	reportnumber = {KEK-TH-2740},
	title = {Monte-{C}arlo studies of the emergent spacetime in the polarized type {IIB} matrix model},
	volume = {135},
	year = {2025}}

@article{Bobev:2025idz,
	archiveprefix = {arXiv},
	author = {Bobev, Nikolay and Mera {\'A}lvarez, Guillermo and Paul, Hynek},
	doi = {10.1007/JHEP07(2025)137},
	eprint = {2503.18770},
	journal = {JHEP},
	pages = {137},
	primaryclass = {hep-th},
	title = {Correlation functions for non-conformal {D$p$}-brane holography},
	volume = {07},
	year = {2025}}

@article{Hartnoll:2025ecj,
	archiveprefix = {arXiv},
	author = {Hartnoll, Sean A. and Liu, Jun},
	doi = {10.21468/SciPostPhys.19.4.099},
	eprint = {2504.06481},
	journal = {SciPost Phys.},
	number = {4},
	pages = {099},
	primaryclass = {hep-th},
	title = {Statistical physics of the polarised {IKKT} matrix model},
	volume = {19},
	year = {2025}}

@article{Ciceri:2025maa,
	archiveprefix = {arXiv},
	author = {Ciceri, Franz and Samtleben, Henning},
	doi = {10.1103/fb8g-b8fd},
	eprint = {2503.08771},
	journal = {Phys. Rev. Lett.},
	number = {6},
	pages = {061601},
	primaryclass = {hep-th},
	title = {Holography for the {I}shibashi-{K}awai-{K}itazawa-{T}suchiya matrix model},
	volume = {135},
	year = {2025}}

@article{Hartnoll:2024csr,
	archiveprefix = {arXiv},
	author = {Hartnoll, Sean A. and Liu, Jun},
	doi = {10.1007/JHEP03(2025)060},
	eprint = {2409.18706},
	journal = {JHEP},
	pages = {060},
	primaryclass = {hep-th},
	title = {The polarised {IKKT} matrix model},
	volume = {03},
	year = {2025}}

@article{Itzhaki:1998dd,
	archiveprefix = {arXiv},
	author = {Itzhaki, Nissan and Maldacena, Juan Martin and Sonnenschein, Jacob and Yankielowicz, Shimon},
	doi = {10.1103/PhysRevD.58.046004},
	eprint = {hep-th/9802042},
	journal = {Phys. Rev. D},
	pages = {046004},
	reportnumber = {TAUP-2474-98, HUTP-98-A003},
	title = {Supergravity and the large ${N}$ limit of theories with sixteen supercharges},
	volume = {58},
	year = {1998}}

@article{Bonelli:2002mb,
	archiveprefix = {arXiv},
	author = {Bonelli, Giulio},
	doi = {10.1088/1126-6708/2002/08/022},
	eprint = {hep-th/0205213},
	journal = {JHEP},
	pages = {022},
	reportnumber = {ULB-TH-02-15},
	title = {Matrix strings in pp wave backgrounds from deformed super {Y}ang-{M}ills theory},
	volume = {08},
	year = {2002}}

@article{Bobev:2018ugk,
	archiveprefix = {arXiv},
	author = {Bobev, Nikolay and Bomans, Pieter and Gautason, Fri\dh{}rik Freyr},
	doi = {10.1007/JHEP08(2018)029},
	eprint = {1805.05338},
	journal = {JHEP},
	pages = {029},
	primaryclass = {hep-th},
	title = {Spherical branes},
	volume = {08},
	year = {2018}}

@article{Bobev:2024gqg,
	archiveprefix = {arXiv},
	author = {Bobev, Nikolay and Bomans, Pieter and Gautason, Fridrik Freyr},
	doi = {10.1007/JHEP01(2025)170},
	eprint = {2410.21376},
	journal = {JHEP},
	pages = {170},
	primaryclass = {hep-th},
	title = {Spherical branes and the {BMN} matrix quantum mechanics},
	volume = {01},
	year = {2025}}

@article{Ciceri:2023bul,
	archiveprefix = {arXiv},
	author = {Ciceri, Franz and Samtleben, Henning},
	doi = {10.1103/PhysRevD.108.106007},
	eprint = {2309.10073},
	journal = {Phys. Rev. D},
	number = {10},
	pages = {106007},
	primaryclass = {hep-th},
	title = {Consistent sphere reductions of gravity to two dimensions},
	volume = {108},
	year = {2023}}

@article{Gubser:1996wt,
	archiveprefix = {arXiv},
	author = {Gubser, Steven S. and Hashimoto, A. and Klebanov, Igor R. and Maldacena, Juan Martin},
	doi = {10.1016/0550-3213(96)00182-4},
	eprint = {hep-th/9601057},
	journal = {Nucl. Phys. B},
	pages = {231--248},
	reportnumber = {PUPT-1586},
	title = {Gravitational lensing by $p$-branes},
	volume = {472},
	year = {1996}}

@article{Ambjorn:2000dx,
	archiveprefix = {arXiv},
	author = {Ambj{\o}rn, Jan and Anagnostopoulos, K. N. and Bietenholz, Wolfgang and Hotta, T. and Nishimura, J.},
	doi = {10.1088/1126-6708/2000/07/011},
	eprint = {hep-th/0005147},
	journal = {JHEP},
	pages = {011},
	reportnumber = {NBI-HE-00-24, NORDITA-2000-50-HE},
	title = {{M}onte {C}arlo studies of the {IIB} matrix model at large {N}},
	volume = {07},
	year = {2000}}

@article{Krauth:1998xh,
	archiveprefix = {arXiv},
	author = {Krauth, Werner and Nicolai, Hermann and Staudacher, Matthias},
	doi = {10.1016/S0370-2693(98)00557-7},
	eprint = {hep-th/9803117},
	journal = {Phys. Lett. B},
	pages = {31--41},
	reportnumber = {AEI-058},
	title = {Monte {C}arlo approach to {M} theory},
	volume = {431},
	year = {1998}}

@article{Bergshoeff:1998ry,
	archiveprefix = {arXiv},
	author = {Bergshoeff, E. and Behrndt, K.},
	doi = {10.1088/0264-9381/15/7/002},
	eprint = {hep-th/9803090},
	journal = {Class. Quant. Grav.},
	pages = {1801--1813},
	reportnumber = {UG-2-98, HUB-EP-14-98},
	title = {D-instantons and asymptotic geometries},
	volume = {15},
	year = {1998}}

@article{Gibbons:1995vg,
	archiveprefix = {arXiv},
	author = {Gibbons, Gary W. and Green, Michael B. and Perry, Malcolm J.},
	doi = {10.1016/0370-2693(95)01565-5},
	eprint = {hep-th/9511080},
	journal = {Phys. Lett. B},
	pages = {37--44},
	reportnumber = {DAMTP-R-95-56},
	title = {Instantons and seven-branes in type {IIB} superstring theory},
	volume = {370},
	year = {1996}}

@article{Ishibashi:1996xs,
	archiveprefix = {arXiv},
	author = {Ishibashi, N. and Kawai, H. and Kitazawa, Y. and Tsuchiya, A.},
	doi = {10.1016/S0550-3213(97)00290-3},
	eprint = {hep-th/9612115},
	journal = {Nucl. Phys. B},
	pages = {467--491},
	reportnumber = {KEK-TH-503},
	title = {A large {N} reduced model as superstring},
	volume = {498},
	year = {1997}}

@article{Ooguri:1998pf,
	archiveprefix = {arXiv},
	author = {Ooguri, Hirosi and Skenderis, Kostas},
	doi = {10.1088/1126-6708/1998/11/013},
	eprint = {hep-th/9810128},
	journal = {JHEP},
	pages = {013},
	reportnumber = {UCB-PTH-98-49A, LBNL-42387, SPIN-1998-01, LBL-42387},
	title = {On the field theory limit of {D}-instantons},
	volume = {11},
	year = {1998}}

@article{Bossard:2023jid,
	archiveprefix = {arXiv},
	author = {Bossard, Guillaume and Ciceri, Franz and Inverso, Gianluca and Kleinschmidt, Axel},
	doi = {10.1007/JHEP01(2024)045},
	eprint = {2309.07233},
	journal = {JHEP},
	pages = {045},
	primaryclass = {hep-th},
	reportnumber = {CPHT-RR062.092023},
	title = {Consistent truncation of eleven-dimensional supergravity on {$S^{8}$\texttimes{} $S^{1}$}},
	volume = {01},
	year = {2024}}

@article{Bianchi:2001kw,
	archiveprefix = {arXiv},
	author = {Bianchi, Massimo and Freedman, Daniel Z. and Skenderis, Kostas},
	doi = {10.1016/S0550-3213(02)00179-7},
	eprint = {hep-th/0112119},
	journal = {Nucl. Phys. B},
	pages = {159--194},
	reportnumber = {MIT-CTP-3166, PUTP-1999, DAMTP-2001-63, ROM2F-2001-30},
	title = {Holographic renormalization},
	volume = {631},
	year = {2002}}

@article{Bossard:2022wvi,
	archiveprefix = {arXiv},
	author = {Bossard, Guillaume and Ciceri, Franz and Inverso, Gianluca and Kleinschmidt, Axel},
	doi = {10.1103/PhysRevLett.129.201602},
	eprint = {2209.02729},
	journal = {Phys. Rev. Lett.},
	number = {20},
	pages = {201602},
	primaryclass = {hep-th},
	reportnumber = {CPHT-RR053.092022},
	title = {Consistent {K}aluza-{K}lein truncations and two-dimensional gauged supergravity},
	volume = {129},
	year = {2022}}

@article{Morales:2004xc,
	archiveprefix = {arXiv},
	author = {Morales, J. F. and Samtleben, H.},
	doi = {10.1016/j.physletb.2004.12.031},
	eprint = {hep-th/0411246},
	journal = {Phys. Lett.},
	pages = {286-293},
	primaryclass = {hep-th},
	reportnumber = {DESY-04-229, CERN-PH-TH-2004-234},
	slaccitation = {%%CITATION = HEP-TH/0411246;%%},
	title = {Higher spin holography for {SYM} in $d$ dimensions},
	volume = {B607},
	year = {2005}}

@article{Nastase:2000tu,
	archiveprefix = {arXiv},
	author = {Nastase, Horatiu and Vaman, Diana},
	doi = {10.1016/S0550-3213(00)00214-5},
	eprint = {hep-th/0002028},
	journal = {Nucl. Phys.},
	pages = {211-236},
	primaryclass = {hep-th},
	reportnumber = {YITP-SB-00-02},
	slaccitation = {%%CITATION = HEP-TH/0002028;%%},
	title = {On the nonlinear {KK} reductions on spheres of supergravity theories},
	volume = {B583},
	year = {2000}}

@article{Cvetic:2000dm,
	archiveprefix = {arXiv},
	author = {Cvetic, Mirjam and Lu, Hong and Pope, C. N.},
	doi = {10.1103/PhysRevD.62.064028},
	eprint = {hep-th/0003286},
	journal = {Phys. Rev.},
	pages = {064028},
	primaryclass = {hep-th},
	reportnumber = {CTP-TAMU-10-00, UPR-881-T},
	slaccitation = {%%CITATION = HEP-TH/0003286;%%},
	title = {Consistent {K}aluza-{K}lein sphere reductions},
	volume = {D62},
	year = {2000}}

@article{deWit:1986oxb,
	author = {de Wit, B. and Nicolai, H.},
	doi = {10.1016/0550-3213(87)90253-7},
	journal = {Nucl.Phys.},
	pages = {211},
	reportnumber = {CERN-TH-4359/86},
	slaccitation = {%%CITATION = NUPHA,B281,211;%%},
	title = {The consistency of the {$S^7$} truncation in {$D=11$} supergravity},
	volume = {B281},
	year = {1987}}

@article{Baguet:2015sma,
	archiveprefix = {arXiv},
	author = {Baguet, Arnaud and Hohm, Olaf and Samtleben, Henning},
	doi = {10.1103/PhysRevD.92.065004},
	eprint = {1506.01385},
	journal = {Phys. Rev.},
	pages = {065004},
	primaryclass = {hep-th},
	reportnumber = {MIT-CTP-4670},
	slaccitation = {%%CITATION = ARXIV:1506.01385;%%},
	title = {Consistent type {IIB} reductions to maximal {5D} supergravity},
	volume = {D92},
	year = {2015}}

@article{Wiseman:2008qa,
	archiveprefix = {arXiv},
	author = {Wiseman, Toby and Withers, Benjamin},
	doi = {10.1088/1126-6708/2008/10/037},
	eprint = {0807.0755},
	journal = {JHEP},
	pages = {037},
	primaryclass = {hep-th},
	slaccitation = {%%CITATION = ARXIV:0807.0755;%%},
	title = {Holographic renormalization for coincident {D$p$}-branes},
	volume = {0810},
	year = {2008}}

@article{Kanitscheider:2008kd,
	archiveprefix = {arXiv},
	author = {Kanitscheider, Ingmar and Skenderis, Kostas and Taylor, Marika},
	doi = {10.1088/1126-6708/2008/09/094},
	eprint = {0807.3324},
	journal = {JHEP},
	pages = {094},
	primaryclass = {hep-th},
	reportnumber = {ITFA-2008-25},
	slaccitation = {%%CITATION = ARXIV:0807.3324;%%},
	title = {{Precision holography for non-conformal branes}},
	volume = {0809},
	year = {2008}}

@article{Boonstra:1998mp,
	author = {Boonstra, H. J. and Skenderis, K. and Townsend, P. K.},
	eprint = {hep-th/9807137},
	journal = {JHEP},
	pages = {003},
	slaccitation = {%%CITATION = HEP-TH 9807137;%%},
	title = {The domain wall/{QFT} correspondence},
	volume = {01},
	year = {1999}}

\end{document}